# Focusing and directional beaming effects of airborne sound through a planar lens with zigzag slits

Kun Tang, [a)]Chunyin Qiu, Jiuyang Lu, Manzhu Ke, and Zhengyou Liu

*Key Laboratory of Artificial Micro- and Nano-structures of Ministry of Education and School of Physics and Technology, Wuhan University, Wuhan 430072, People's Republic of China*

**Abstract:** Based on the Huygens-Fresnel principle we design a planar lens to efficiently realize the interconversion of the point-like source and Gaussian beam in the air ambience. The lens is constructed by a planar plate drilled elaborately with a nonuniform array of zigzag slits, where the slit exits act as subwavelength-sized secondary sources carrying desired sound responses. The experiments operated at audible regime agree well with the theoretical predictions. This compact device could be useful in daily life applications, such as for medical and detection purposes.

[a)]Author to whom correspondence should be addressed. Emails: cyqiu@whu.edu.cn

Highly efficient transmitted devices for airborne sound have been challenging for a long time. A critical factor is the extreme impedance contrast between air and natural solids, which strongly suppresses the coupling efficiency of the acoustic devices. This problem is considerably relaxed by the recent development of the artificial structures [1-11], such as sonic crystals and acoustic metamaterials. By employing these artificial materials, numerous transmitted devices have been proposed, e.g. flat lenses with imaging effects [2-7]. The improvement of the transparency in such artificial materials can be mostly attributed to the existence of air channels for direct sound propagation.

Focusing of the airborne sound has always been a hot topic in the field of acoustic artificial structures. In previous studies, curved lenses built by uniform structure units [1,12-15] or planar lenses constructed from spatially gradient building blocks [16-20] are frequently proposed. In the former case the focusing effect stems from the curved shape of the lens, whereas in the latter case the focusing effect depends on the gradient distribution of the local refraction index. Therefore, in these designs the geometric acoustics (i.e. ray theory) is usually implied, which consider mainly the (refraction-index-weighted) acoustic path length and ignore the multi-reflections at the device's interfaces. This simplification in turn influences the predefined amplitude and phase emitting from the lens and, as a result, degrades the quality of focusing.

Here we propose a focusing lens that is built by a planar plate drilled with zigzag slits (ZSs). Our theoretical and experimental results consistently manifest a high-quality focusing effect through such a lens with planar-like surface shape. The design starts from a reverse process according to the Huygens-Fresnel principle: each slit is elaborately constructed to enable its exit with a wave response equivalent to that emitting from the prescribed focusing point. Similar design strategies have been widely employed to design optical metasurfaces with great flexibility in tailoring wavefronts [21-24], such as redirecting the propagation of incident waves. Recently, analogous studies have been extended to sound waves [25-28]. However, so far there is no such focusing lens realized for transmitted sound waves. It is of interest that the planar lens presented here also exhibits an excellent conversion from a subwavelength-sized point source to a highly directional Gaussian beam. Throughout the paper, all full-wave simulations are performed by the commercial finite-element solver (COMSOL Multiphysics), where the solid (used to construct the lens) is modeled as acoustically rigid with respect to air.

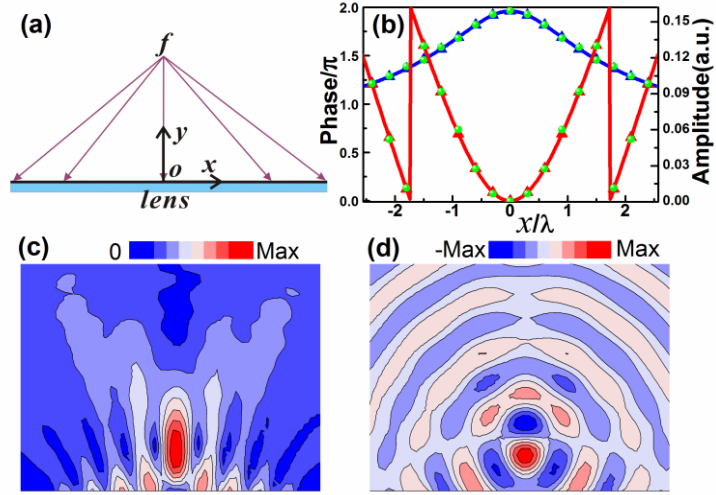

FIG. 1. (a) Schematic of the reverse design for the focusing lens, where $f$ is the desired focusing point. (b) The retrived continuous amplitude (blue line) and phase (red line) distributions at the outgoing surface of the lens, where the descrete blue and red triangles are utilized to demonstate the effectiveness of the reverse design in (c) and (d), and the green dots indicate the comparing data extracted from the real design in Fig. 2. (c) The amplitude field distribution superposed from the discrete point sources arranged on the outgoing surface. (d) The corresponding instant field pattern.

Here we consider the acoustic focusing effect for a plane wave transmitting through a planar lens, where the desired focusing point is set $d = \lambda$ away from the outgoing surface, with $\lambda$ being the wavelength in air. According to the Huygens-Fresnel principle, the sound field transmitted through a device can be viewed as a superposition of the secondary waves emitting from all points on the device's outgoing surface. As shown in Fig. 1(a), in the reverse design the desired focusing point $f$ is placed with a point source, which is characterized by the 0-order of Hankel function of the first kind, i.e. $H_0^{(1)}(kr)$. Here $r$ is the distance from the $f$ point and $k$ is the wavevector in free space. In this way the amplitude and the phase on the outgoing surface of the lens can be retrieved from the emitting wave of the point source, as depicted in Fig. 1(b). Note that in the practical design, only finite and discrete (instead of infinite and continuous) points carrying desired sound responses can be mimicked. To demonstrate the effectiveness of the reverse design, a total of 17 discrete points, each assigned with specific amplitude and phase responses [see triangles in Fig. 1(b)], are located on the outgoing surface of the virtual lens with a spatial step of $0.3\lambda$. As shown in Figs. 1(c) and 1(d), the superposing amplitude and temporal fields show an excellent focusing effect, although the emission surface has a length of $\sim 5\lambda$ only.

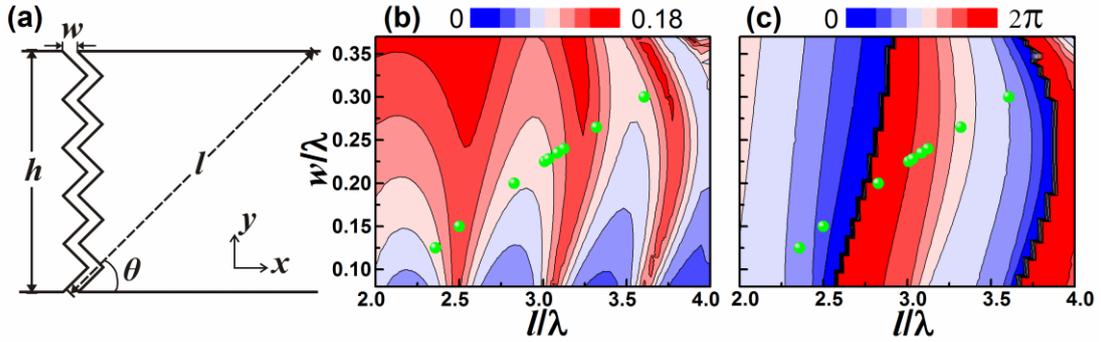

FIG. 2. (a) Schematic of a zigzag slit perforated in a planar plate of thickness $h = 2\lambda$, where the width $w$ and the total length $l = h/\sin\theta$ of the slit can be adjusted. (b) The far-field amplitude reponse (in arbitary units) varied with $w$ and $l$, excited by a plane wave impinged onto the single slit system along the $+y$ direction. (c) The corrseponding phase shift across the slit. Here the green dots indicate the slit parameters selected in the real design.

Below we demonstrate that a nonuniform array of ZSs can minic the above secondary sources retrived from the reverse design. Recently, the ZS units have been extensively employed to design acoustic metamaterials [9-11,19,20] and metasurfaces [25-28] due to the flexibility in tailoring the propagation length of sound. As illustrated in Fig. 2(a), the ZS is drilled through a planar plate of height $h = 2\lambda$. To reduce the number of the adjustable geometric parameters, each slit is regularly zigzagged with ten segments and characterized by two parameters, i.e. the slit width $w$ and the inclining angle $\theta$, where the latter gives equivalently the total slit length $l = h/\sin\theta$. To find appropriate ZSs that fulfill the required sound responses, a plane wave is normally incident upon the planar plate perforated with a single slit, and the amplitude and phase in the outgoing far-field are collected. Since the ZS has a width of deep subwavelength, the sound wave emitting from the slit exit can be safely regarded as a point-like source. From the far-field sound response, it is easy to derive the amplitude and phase shift at the exit. Intuitively, the wave amplitude is expected to be tailored mostly by the slit width $w$, and the phase shift should be simply proportional to the elongated total length $l$. However, this prediction is not so precise considering the interference among the waves traveling back and forth inside the ZS (due to the impedance mismatch at the slit exits). In Figs. 2(b) and 2(c), we present respectively the simulated amplitude and phase responses for the sound emitting from the exit, varying with the geometric parameters $w$ and $l$ (instead of $\theta$). It is observed that the sound responses cover a wide range of values. In particular, the phase shift runs over the complete $2\pi$ span, where the values $\sim n\pi$ (associated with maximal amplitudes) correspond to the well-known Fabry-Perot resonances. From the Figs. 2(b) and 2(c) several ZSs (green dots) can be picked out for the real design, associated with geometric details listed in Table 1. The corresponding sound responses (green dots) are also presented in Fig. 1(b), which fit those ideal amplitudes (blue triangles) and phases (red triangles) very well. As mentioned above, these ZS units are orderly arranged with a step of $0.3\lambda$ and form a planar lens of length $\sim 5\lambda$. In Fig. 3(a) we present the amplitude profile for a Gaussian beam of width $\sim 7.5\lambda$ impinging normally onto the designed lens. It is observed that a well-defined focusing spot emerges with a focal length $\sim 1\lambda$, in good agreement with that displayed in Fig. 1(c) attained from the simple model.

| unit # | 1 (17) | 2 (16) | 3 (15) | 4 (14) | 5 (13) | 6 (12) | 7 (11) | 8 (10) | 9 |
|---|---|---|---|---|---|---|---|---|---|
| $l$ | 3.01 | 3.12 | 3.61 | 2.82 | 3.08 | 3.32 | 2.36 | 2.5 | 3.03 |
| $w$ | 0.225 | 0.24 | 0.30 | 0.20 | 0.235 | 0.265 | 0.125 | 0.15 | 0.228 |

Table 1. Geometric parameters (scaled by λ) used for the 17 zigzag slits.

Now we present the experimental validation for the above predicted focusing effect. The experimental set up is briefly described as follows. With carefully considering the multi-scale nature of the whole system, the working wavelength in the experiment is chosen as $\lambda = 8.5$cm (i.e. at frequency 4000Hz), which gives all geometry parameters (which are scaled by $\lambda$). The sample is fabricated by sculpturing a Plexiglass plate (of thickness 1.2cm), which behaves acoustically rigid with respect to air. It is tightly sandwiched between a laboratory table and a covering Plexiglass plate, where the parallel gap in between supports only the fundamental wave-guiding mode to guarantee a two-dimensional sound propagation. To reduce the undesired reflection from the free space, absorbers are placed at the exits of the waveguide. To mimic the incident wave used in the full-wave simulation, a Gaussian beam of width $\sim 7.5\lambda$ has been produced by a subwavelength-sized microphone together with a parabolic concave-mirror [29]. The field distribution behind the sample is measured by a couple of identical microphones (of diameter $\sim 0.1\lambda$, B&K Type 4187): one is fixed for phase reference, and the other is movable for field scanning. In Fig. 3(b) we present the measured sound field distribution, which exhibits a high-quality focusing spot with focal length $\sim 1\lambda$. To make a more quantitative comparison between the experimental and numerical data, in Figs. 3(c) and 3(d) we present the intensity distributions along the horizontal and vertical lines drawn in Figs. 3(a) and 3(b), which go through the focal spots. It is observed that the experimental results (circles) agree excellently well with the full-wave simulations (lines). Both the measured and simulated focal spots can be characterized by the widths of the half-intensities, i.e. $\sim 0.4\lambda$ along the *x* direction and $\sim 1.2\lambda$ along the *y* direction. Compared with most of the previous works [1,12-20], the design here shows the superiority for the weaker elongation of the focal spot along the propagation direction.

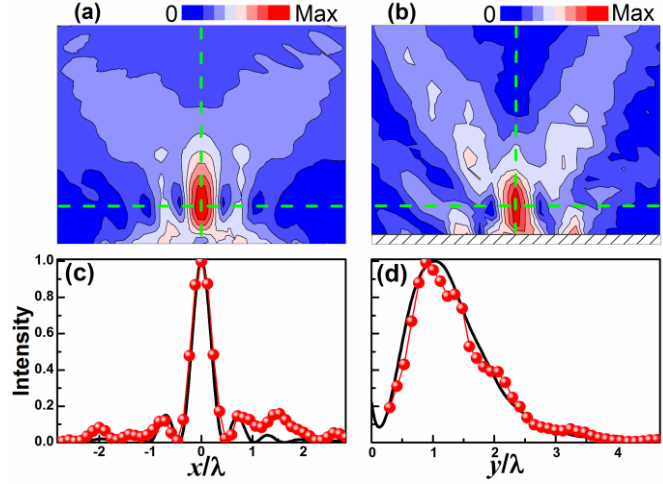

FIG. 3. (a) The numerical amplitude profile for the focusing effect, obtained by a Gaussian beam incident normally upon the sample. (b) The corrsponding experimental result, where the sound field in the shadow region (very close to the lens) is not scanned. (c) Pressure intensities (normalized by the maxium values) distributed at the horizontal lines marked in (a) and (b), where the black solid line and the red circle denote the numerical and experimental data, respectively. (d) The same as (c), but for the vertical lines.

So far we have demonstrated the efficient focusing effect for a Gaussian beam transmitted through the carefully designed planar lens. Below we manifest that this planar lens can also transform the point-like source into a well-collimated Gaussian beam. In Figs. 4(a) and 4(b), we present the numerical field patterns excited by a point source with $\sim 1\lambda$ away from the lens surface (i.e. positioned at the aforementioned focusing spot). Both the amplitude and temporal fields transmitting through the planar lens exhibit an excellent collimating behavior. In Figs. 4(c) and 4(d) we present the corresponding experimental results, which are excited by the sound wave guided from a subwavelength-sized tube (with a diameter $\sim 0.1\lambda$). It is observed that the measured data coincide reasonably well with the full-wave simulations. Note that in previous works, the highly directional sound beaming effect has been realized with using sonic crystals [30-34] and grating structures [35-39]. The former is frequently accompanied with the excitation of specific states in the crystals, e.g. the cavity states [30-32] and band-edge states [33], whereas the latter is often connected with the resonant excitation of the structure-induced surface acoustic waves [35-37]. Different from those solutions invovling periodic structures, the design here considers a nonuniform structure and thus becomes more flexible, such as in tailoring the direction of the outgoing beam.

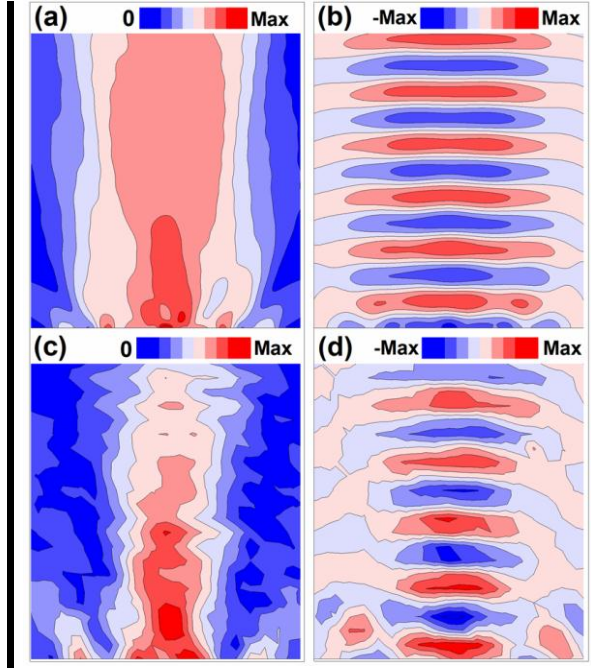

FIG. 4. The directional beaming effect for a point-like source transmitted through the planar lens, where (a) and (b) correspond to the simulated amplitude and temporal field distributions, and (c) and (d) represent the corresponding experimental comparisons, respectively.

In summary, a high-quality focusing effect has been demonstrated for a planar lens arranged by a finite array of zigzag slits. It follows a reverse design based on the Huygens-Fresnel principle, in which both the amplitude and phase responses are fully taken into account. It is of interest that such a planar lens can also transform a point-like source into a well-collimated Gaussian beam. The performance of the wavefront conversion has been further validated by real experiments. In principle, a three-dimension version of the planar lens can be directly designed with using zigzag hole-arrays. The proposed design strategy can also be flexibly extended to the other wavefront manipulations.

**Acknowledgements**

This work is supported by the National Natural Science Foundation of China (Grant

Nos. 11174225, 11004155, 11374233, and J1210061); National Basic Research Program of China (Grant No. 2015CB755500); Open Foundation from State Key Laboratory of Applied Optics of China; and the Program for New Century Excellent Talents (NCET-11-0398).